\documentclass{ws-procs975x65}

\begin{document}

\title{Covariant Description of the Inhomogeneous Mixmaster Chaos}

\author{R. Benini$^{12\dag}$ and G. Montani$^{23\ddag}$}

\address{$^1$Dipartimento di Fisica - Universit\`a di Bologna and INFN\\ Sezione di Bologna,
via Irnerio 46, 40126 Bologna, Italy\\
$^2$ICRA---International Center for Relativistic Astrophysics  
c/o Dipartimento di Fisica (G9) Universit\`a di Roma ``La Sapienza'',
Piazza A.Moro 5 00185 Roma, Italy\\
$^3$ENEA C.R. Frascati (U.T.S. Fusione),\\ via Enrico Fermi 45, 00044
Frascati, Rome, Italy\\
$^\dag$\email{riccardo.benini@icra.it}
$^\ddag$\email{montani@icra.it}}

\begin{abstract}
We outline the covariant nature of the chaos characterizing the generic cosmological solution near the initial singularity. Our analysis is based on a "gauge" independent ADM-reduction of the dynamics to the physical degrees of freedom, and shows that the dynamics is isomorphic point by point in space to a billiard on a Lobachevsky plane. The Jacobi metric associated to the geodesic flow is constructed and a non-zero Lyapunov exponent is explicitly calculated. The chaos covariance emerges from the independence of the form of the lapse function and the shift vector.
\end{abstract}

%\keywords{Style file; \LaTeX; Proceedings; World Scientific Publishing.}

\bodymatter

In recent years, the chaoticity of the homogeneous Mixmaster model has been widely studied in the literature (see \cite{KirillovMontani1997PRD,CornishLevin1997PRD,ImponenteMontani2001PRD, Montani2001CQG}) in view of understanding the features of its covariant nature.
 Two convincing arguments, appeared in \cite{CornishLevin1997PRD, ImponenteMontani2001PRD}, support the idea that the Mixmaster chaos (described by the invariant measure introduced in \cite{ChernoffBarrow1983PRL, KirillovMontani1997PRD}) remains valid in any system of coordinates.\\
 The main issue of the present work is to show that the property of space-time covariance can be extended to the inhomogeneous Mixmaster model.\\
A generic cosmological solution is represented by a gravitational field having available all its degrees of 
freedom and, therefore, allowing to specify a generic Cauchy problem.
In the Arnowitt-Deser-Misner (ADM) formalism, the metric tensor
corresponding to such a generic model takes the form
\begin{equation}
	d\Gamma^2=N^2 dt^2-\gamma_{\alpha\beta}(dx^\alpha+N^\alpha dt)(dx^\beta+N^\beta dt)
\end{equation}
where $N$ and $N^\alpha$ denote (respectively) the lapse function and the shift-vector, $\gamma_{\alpha\beta}$ the 3-metric tensor of the spatial hyper-surfaces $\Sigma^3$ for which $t=const$, being\cite{BeniniMontani2004PRD}
\begin{equation}
\label{parametrizzazione della metrica}
	\gamma_{\alpha\beta}=e^{q_a}\delta_{ad}O^a_b O^d_c \partial_\alpha y^b \partial_\beta y^c\ \ \ 
	a,b,c,d,\alpha,\beta=1,2,3;
\end{equation} 
 $q^a=q^a(x,t)$ and $y^b=y^b(x,t)$ are six scalar functions, and $O^a_b=O^a_b(x)$ a $SO(3)$ matrix.\
By the metric tensor (\ref{parametrizzazione della metrica}), the action for the gravitational field is
\begin{equation}
\label{azione standard}
	S=\int_{\Sigma^{(3)}\times\Re}dt d^3 x\left(p_a\partial_t q^a+\Pi_d\partial_t y^d -NH-N^\alpha H_\alpha\right)\,,
\end{equation}
\begin{equation}	
	%\begin{subequation}
	\label{vincoli hamiltoniani} 
	H=\frac{1}{ \sqrt \gamma}[\sum_a (p_a)^2-\frac{1}{2}(\sum_b p_b)^2-\gamma ^{(3)}R]
\end{equation}	
\begin{equation}
\label{vincoli hamiltoniani2}	 
	 H_\alpha=\Pi_c \partial_\alpha y^c +p_a \partial_\alpha q^a +2p_a(O^{-1})^b_a\partial_\alpha O^a_b;
	%\end{subequation}
\end{equation}
in (\ref{vincoli hamiltoniani}) and (\ref{vincoli hamiltoniani2}) $p_a$ and $\Pi_d$ are the conjugate momenta to the variables $q^a$ and $y^b$ respectively, and the $^{(3)}R$ is the Ricci 3-scalar which plays the role of a potential term.
We use the Hamiltonian constraints $H=H_\alpha=0$ to reduce the dynamics to the physical degrees of freedom; the super-momentum constraints can be diagonalized and explicitly solved by choosing the function $y^a$ as special coordinates:
\begin{equation}
	\Pi_b=-p_a\frac{\partial q^a}{\partial y^b}-2p_a(O^{-1})^c_a\frac{\partial O^a_c}{ \partial y^b}\,.
\end{equation}
\begin{equation}
	\label{finale non approssimata}
	S=\int_{\Sigma^{(3)}\times\Re}d\eta d^3 y \left(p_a\partial_\eta q^a+2p_a(O^{-1})^c_a\partial_\eta O^a_c-NH\right)\,.
\end{equation} 
The potential term appearing in the super-Hamiltonian  behaves as an infinite potential wall as the determinant of the 3-metric  goes to zero, and we can model it as follows 
\begin{equation}
\label{U}
	U=\sqrt{\gamma} ^{(3)}R=\sum_a{\Theta(Q_a)} \hspace{1.0cm} 	\Theta(x)=\begin{cases}
	+\infty\ &  $if$\  x<0,\cr 0\ &  $if$\  x>0.\cr
	\end{cases}
\end{equation}
where the $Q_a$'s are called {\it anisotropy parameters}. By (\ref{U}) the Universe dynamics evolves independently in each space point and the point-Universe can move  within a dynamically-closed domain $\Gamma_Q$ only\cite{BeniniMontani2004PRD} (see figure (\ref{fig:cuspidi2 cap5})).
Since in $\Gamma_Q$ the potential $U$ asymptotically vanishes, near the singularity we have $\partial p_a/ \partial\eta=0$; then the term $2p_a(O^{-1})^c_a\partial_\eta O^a_c$ in (\ref{finale non approssimata}) behaves as an exact time-derivative that can be ruled out of the variational principle.\\
The ADM reduction is completed by introducing the so-called Misner-Chitr\'e like $(\tau, \xi,\theta)$ variables\cite{ImponenteMontani2001PRD}, in  terms of which  the anisotropy parameters $Q_a$ become $\tau$ independent:
\begin{figure}[ht]
 \begin{minipage}[b]{0.5\textwidth}
 \begin{flushleft}
\begin{equation}\nonumber
Q_1=\frac{1}{ 3}-\frac{\sqrt{\xi^{2}-1}}{ 3\xi}(\cos\theta+\sqrt{3}\sin\theta)
\end{equation}
\begin{equation}
\label{parametri di anisotropia32}
        Q_2=\frac{1}{ 3}-\frac{\sqrt{\xi^{2}-1}}{ 3\xi}(\cos\theta-\sqrt{3}\sin\theta)
\end{equation}
\begin{equation}\nonumber
Q_3=\frac{1}{ 3}+\frac{2\sqrt{\xi^{2}-1}}{ 3\xi}\cos\theta
\end{equation}
\end{flushleft}
\end{minipage}%
\begin{minipage}[b]{0.5\textwidth}
\begin{center}
\includegraphics[width=5cm]{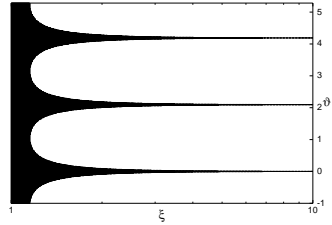} 
\caption{ $\Gamma_Q(\xi,\theta)$\label{fig:cuspidi2 cap5}} 
\end{center}    
\end{minipage}
\end{figure}
When expressed in terms of such variables the super-Hamiltonian constraint can be solved in the domain $\Gamma_Q$ according to the ADM prescription:
\begin{equation}
	\label{hamiltoniana ADM}
	-p_\tau\equiv\epsilon=\sqrt{(\xi^2-1)p_\xi^2+\frac{p_\theta^2}{\xi^2-1}}
\end{equation}
\begin{equation}
	\label{hamiltoniana ridotta}
	\delta S_{\Gamma_Q}=\delta\int d\eta d^3 y (p_\xi\partial_\eta\xi+p_\theta\partial_\eta\theta-\epsilon\partial_\eta\tau)=0\,.
\end{equation}
By virtue of the asymptotic limit (\ref{U}) and the Hamilton equations associated with (\ref{hamiltoniana ridotta}) it follows that $\epsilon$ is a constant of motion, i.e. $d\epsilon/ d\eta=\partial\epsilon/\partial\eta=0\Rightarrow \epsilon=E(y^a)$; furthermore $\epsilon\partial_\eta\tau$ behaves as an exact time derivative.\\
This dynamical scheme allows us to construct the Jacobi metric corresponding to the dynamical flow, and the line element reads
\begin{equation}
\label{elemento di linea}
	ds^2=E^{2}(y^a)\left(\frac{d\xi^{2}}{ (\xi^{2}-1)}+(\xi^{2}-1)d\theta^{2}\right)\,.
\end{equation}
Here the space coordinates behaves like external parameters since the evolution is spatially uncorrelated.
The Ricci scalar takes the value $R=-2/E^2$, so that (\ref{elemento di linea}) describes a two-dimensional Lobachevsky space; the role of the potential wall (\ref{U}) consists of cutting a closed domain $\Gamma_Q$ on such a negative curved surface.  \\
In this scheme the Lyapunov exponent associated to the geodesic deviation along the direction orthonormal to the geodesic 4-velocity can be evaluated, and it results to be equal to 
\begin{equation}
	\label{esponente di lyapunov}
	\lambda(y^a)=\frac{1}{ \epsilon(y^a)}>0\,.
\end{equation}
Hence the chaotic behavior of the inhomogeneous Mixmaster model can be described in a generic gauge ({\it i.e.} without assigning the form of the lapse function and of the shift vector) as soon as Misner-Chitr\'e like variables are adopted. The value (\ref{esponente di lyapunov}) is a positive definite function of the space point, making this  calculation extendible point by point to the whole space.

\end{document}